\documentclass[letterpaper, 10 pt, conference]{ieeeconf}  

\IEEEoverridecommandlockouts                              
\usepackage{amsmath} 
\usepackage{array}
\usepackage{balance}
\usepackage{rotating}
 \usepackage[titlenumbered,ruled]{algorithm2e}
\usepackage{listings}
\usepackage{amsmath}

\usepackage{makecell}
\usepackage{longtable}
\usepackage{array}
\usepackage{amssymb}
\usepackage{setspace}
\usepackage{eurosym}

\usepackage{color}
\usepackage[usenames,dvipsnames,svgnames,table]{xcolor}

\usepackage[font=small,justification=centering]{caption}

\renewcommand{\baselinestretch}{0.93}

\title{\LARGE \bf
An XML-based Factory Description Language for Smart Manufacturing Plants in Industry 4.0}

\author{Shuai Zhao, Piotr Dziurzanski and Leandro Soares Indrusiak\\
University of York, Deramore Lane,
York, YO10 5GH, UK.\\
{\tt \{firstname.lastname\}@york.ac.uk}\\
}

\begin{document}

\maketitle
\thispagestyle{empty}
\pagestyle{empty}

\begin{abstract}
Industry 4.0 revolution concerns the digital transformation of manufacturing
and promises to answer the ever-increasing demand of product customisation
and manufacturing flexibility while incurring low costs. To perform the required factory reconfiguration, a computationally demanding optimisation process has to be executed to find favourable solutions in a relatively short time. While previous research focused on planning and scheduling of smart factories based on cloud-based optimisation, little attention has been paid to effective approaches to describe the targeted factory, the required products and the production processes. However, these matters are fundamental for the optimisation engine to be correctly and efficiently performed.
This paper presents an XML-based factory modelling language to effectively describe the above data for a given factory and commodity order and to provide a convenient interface for altering the input information. Finally, two real-world manufacturing plants are provided to illustrate the feasibility of the proposed description language.
\end{abstract}

\section{Introduction}
To remain competitive in the industry 4.0 era, smart factories need to be adapted to small batches and highly customised manufacturing \cite{Wang2016}. These new conditions require dynamic adaptivity of the factory with regards to process planning and scheduling, governed by enterprise resource planning (ERP) and manufacturing execution systems (MES) connected to smart devices (things) \cite{Wan19}. An integrated process planning and scheduling is triggered after receiving a new manufacturing order or when a smart factory state change has been detected, caused by e.g. a thing failure \cite{Dziurzanski2018b}. Typically, the current factory state is inferred based on the values read from individual things and sent to an optimisation engine together with the new manufacturing order to be allocated and scheduled \cite{Alsafi2010}. The final solution is then sent to the users and/or applied automatically to things \cite{Wan19,Givehchi17}.

Several works have been proposed aiming at dynamic planning and scheduling of smart manufacturing plants based on multi-criteria optimisations~\cite{Dziurzanski18,Dziurzanski2018b,Dziurzanski19,Zhao19}. However, these works are based on the assumption that the optimisation objectives, required products, available machines, the potential co-relations and synchronisations during the manufacturing process are correctly and effectively described, and are provided to the methods proposed in above works. However, such a factory description (modelling) approach has not been widely discussed.

In this paper, an XML-based language for factory description, named Factory Description Language (FDL), is proposed. This language is expressive sufficiently to describe manufacturing plants considered in this and above papers, and a wide range of other business cases. The motivation behind FDL is to make the description and modelling process human-readable and editable, so that engineers without a high level of expertise in factory modelling can relatively easily define the problem to be optimised, including a plant architecture, products, production processes etc.

This paper is organised as follows. Section~\ref{sec:model} describes the studied manufacturing plants in this work. Section~\ref{sec:fdl} presents the proposed factory description language. Section~\ref{sec:usecase} provides two use-case examples with the adoption of FDL. Finally, Section~\ref{sec:conclusion} draws the conclusion.

\section{Factory Model and Characterisation} \label{sec:model}

In this paper, two typical types of manufacturing plants are considered related to process manufacturing and discrete manufacturing. In discrete manufacturing, products are distinct and easily countable, whereas products under process manufacturing are usually manufactured in bulk and the ingredients are combined as specified in a recipe.

For orders issued by customers, an optimisation is performed towards the planning and scheduling for producing the specified amount of commodities, based on more than one objectives. In the targeted plant, there exists a set of devices, which can be used for manufacturing the required commodities.
The optimisation process concerns mainly the decisions of the time and device that an order should be executed on, to achieve the pre-defined objectives.

For a given order (a certain amount of a given commodity), a pre-defined set of operations are required to be performed for manufacturing. As described above,  such operations may involve several types of device in an explicit execution order. We refer to such a set of devices as production line hereafter.
In addition, a commodity can be produced under a given device with different operating modes, where different modes require varied computation time, energy and monetary cost.

Additionally, when a given device finishes its current production and before it starts producing the next commodity, the device usually needs to be cleaned up and prepare the raw materials for the next commodity, especially when the subsequent commodities are of different types. This phase is referred to as the \textit{sequence-dependent setup} phase, which imposes extra time and cost. The sequence-dependent setup phase is also considered into the FDL and the optimisation engines proposed in~\cite{Dziurzanski2018b,Zhao19}.

\section{Factory Description Language} \label{sec:fdl}

The elements (tags) of FDL directly correspond with the principles described earlier in Section~\ref{sec:model}. Below, we discuss the most important components of the proposed factory description language.

Element \textit{objectives} includes a set of elements named \textit{objective}, where each \textit{objective} represents one objective of the optimisation problem. These objectives will be passed directly into the multi-criteria optimisation engine as the optimisation objectives. Below, the template for describing the objectives is presented.

\lstset{language=XML,tabsize=2,
    commentstyle=\color{OliveGreen},
    stringstyle=\color{red},
    numbers=none,
    numberstyle=\tiny,
    numbersep=5pt,
    breaklines=true,
    showstringspaces=false,
    basicstyle=\footnotesize,
    emph={name,availability,energyConsumption,monetaryCost,start,end},
    xleftmargin=0.1cm,
    columns=flexible,
    emphstyle={\color{blue}}
    }
\begin{lstlisting}
<objectives>
    <objective name="objective1" />
    <objective name="objective2" />
    <objective name="objective3" />
</objectives>
\end{lstlisting}

Element \textit{processingDevices} includes a set of elements named \textit{processingDevice}, representing all processing resources (e.g. machines) in a plant. A \textit{processingDevice} element requires the name attribute. As a resource can operate in a number of various operation modes, a \textit{processingDevice} element includes the nested \textit{modes} element, which in turn includes a set of \textit{mode} elements with the mandatory \textit{name argument}. Each resource has to include at least one mode.

\lstset{language=XML,tabsize=2,
    commentstyle=\color{OliveGreen},
    stringstyle=\color{red},
    numbers=none,
    numberstyle=\tiny,
    numbersep=5pt,
    breaklines=true,
    showstringspaces=false,
    basicstyle=\footnotesize,
    emph={name,availability,energyConsumption,monetaryCost,start,end},
    xleftmargin=0.1cm,
    columns=flexible,
    emphstyle={\color{blue}}
    }
\begin{lstlisting}
<processingDevices>
    <processingDevice name= "device1">
    	<modes>
    		<mode name= "mode1"/>
    		<mode name= "mode2"/>
    		<mode name= "mode3"/>
    	</modes>
    </processingDevice>
</processingDevices>
\end{lstlisting}

The \textit{productionLines} element describes all production lines in a factory, introduced as nested \textit{productionLine} elements. The name attribute in the \textit{productionLine} element is mandatory. The \textit{productionLine} element includes a nested \textit{productionLineProcessingDevices} element, which in turn includes nested \textit{productionLineProcessingDevice} elements. Each \textit{productionLineProcessingDevice} start-tag includes two attributes, \textit{order} and \textit{name}. The former attribute values are consecutive numbers that identify the resource order in a production line, whereas the latter attribute values have to be equal to the resource names introduced in element \textit{processingDevice}. Each production line is linear and thus each possible split of processing results in creating a new production line, from the production line source to its sink. In the example below, the two production lines starts with the same resource (\textit{Scale}), but as two routes are possible starting from \textit{Converyor1} or \textit{Conveyor2}, two \textit{productionLine} elements starting from the \textit{Scale} resource are generated.

\lstset{language=XML,tabsize=2,
    commentstyle=\color{OliveGreen},
    stringstyle=\color{red},
    numbers=none,
    numberstyle=\tiny,
    numbersep=5pt,
    breaklines=true,
    showstringspaces=false,
    basicstyle=\footnotesize,
    emph={name,availability,energyConsumption,monetaryCost,start,end},
    xleftmargin=0.1cm,
    columns=flexible,
    emphstyle={\color{blue}}
    }
\begin{lstlisting}
<productionLines>
	<productionLine name="ProductionLine1">
		<productionLineProcessingDevices>
			<productionLineProcessingDevice order="1" name="device1"/>
			<productionLineProcessingDevice order="2" name="device2"/>
			<productionLineProcessingDevice order="3" name="device3"/>
		</productionLineProcessingDevices>
</productionLines>
\end{lstlisting}

Element \textit{productionProcesses} includes a set of production processes that need to be scheduled in the considered plant. Each \textit{productionProcess} element, nested in \textit{productionProcesses}, includes the mandatory \textit{name} attribute and one or more alternative sets of \textit{subprocesses} leading to manufacturing a certain commodity.
Each \textit{subprocess} element requires the \textit{name} attribute and a set of nested \textit{subprocessProcessingDevice} elements. Unique names of subprocesses are required to refer to them unambiguously from other elements, e.g. \textit{sequenceDependentSetup} (explained later). If more than one \textit{subprocessProcessingDevice} elements are provided, they are treated as alternative ones and being capable of producing the same commodity.

In the \textit{subprocessProcessingDevices} element, all processing devices that have to be allocated simultaneously to execute the given subprocess are listed with elements \textit{subprocessProcessingDevice}. The mandatory argument of this tag is \textit{processingDeviceName}, whose value shall be found in the \textit{processingDevice} element described earlier. Then \textit{subprocessProcessingDevicesMode} elements follow with the mandatory \textit{modeName} attribute whose value shall be listed into the corresponding \textit{processingDevice} element, as described earlier. The \textit{subprocessProcessingDevicesMode} element includes at least one of the three elements: \textit{processingTime}, \textit{energyConsumption} and \textit{monetaryCost}. These three elements specify the corresponding numeric costs of using the particular processing device in the particular mode and as such can be later used to define a fitness function of a factory scheduling.
The usage of these elements is demonstrated in the following example.

\lstset{language=XML,tabsize=2,
    commentstyle=\color{OliveGreen},
    stringstyle=\color{red},
    numbers=none,
    numberstyle=\tiny,
    numbersep=5pt,
    breaklines=true,
    showstringspaces=false,
    basicstyle=\footnotesize,
    emph={name,availability,processingTime,energyConsumption,monetaryCost,start,end},
    xleftmargin=0.1cm,
    columns=flexible,
    emphstyle={\color{blue}}
    }
\begin{lstlisting}
<productionProcesses>
	<productionProcess name="production1">
		<subprocesses>
			<subprocess name="production1Task1">
				<subprocessProcessingDevices>
					<subprocessProcessingDevice processingDeviceName="device1">
						<subprocessProcessingDeviceMode modeName="mode1">
							<processingTime>x1</processingTime>
							<energyConsumption>y1</energyConsumption>
							<monetaryCost>z1</monetaryCost>
						</subprocessProcessingDeviceMode>
					</subprocessProcessingDevice>
					<subprocessProcessingDevice processingDeviceName ="device1">
						<subprocessProcessingDeviceMode modeName="mode2">
							<processingTime>x2</processingTime>
							<energyConsumption>y2</energyConsumption>
							<monetaryCost>z2</monetaryCost>
						</subprocessProcessingDeviceMode>
					</subprocessProcessingDevice>
				</subprocessProcessingDevices>
			</subprocess>
		</subprocesses>
	</productionProcess>
</productionProcesses>
\end{lstlisting}

Another element that is mandatory in a \textit{productionProcess} element, as long as that element includes more than one \textit{subprocess} element, is \textit{subprocessRelations}, using \textit{subprocessRelation} to describe relations between subprocesses in the considered \textit{productionProcess}. Three arguments are mandatory: \textit{source} and \textit{destination} require a proper name of subprocess introduced in the considered \textit{productionProcess}, whereas \textit{allensOperator} requires any relation from the interval Allen's algebra that describes the temporal relation between the source and the destination. The following \textit{allensOperator} values are possible: \textit{LT} for source earlier than destination, \textit{S} for source since destination, \textit{F} for finish destination, \textit{EQ} for source equal to destination, \textit{O} for source overlapping destination, \textit{M} for source meeting destination and \textit{D} for source during destination.
Below, an FDL template for describing subprocess relations is presented.

\lstset{language=XML,tabsize=2,
    commentstyle=\color{OliveGreen},
    stringstyle=\color{red},
    numbers=none,
    numberstyle=\tiny,
    numbersep=5pt,
    breaklines=true,
    showstringspaces=false,
    basicstyle=\footnotesize,
    emph={name,availability,source,destination,allensOperator,start,end},
    xleftmargin=0.1cm,
    columns=flexible,
    emphstyle={\color{blue}}
    }
\begin{lstlisting}
<subprocessRelations>
	<subprocessRelation source="Task1" destination="Task2" allensOperator="M"/>
	<subprocessRelation source="Task2" destination="Task3" allensOperator="M"/>
	<subprocessRelation source="Task3" destination="Task4" allensOperator="M"/>
	<subprocessRelation source="Task4" destination="Task5" allensOperator="M"/>
</subprocessRelations>
\end{lstlisting}

Element \textit{sequenceDependentSetup} determines extra costs when two certain subprocesses, specified with attributes \textit{source} and \textit{destination}, are performed subsequently using the same processing device, specified with attribute \textit{processingDevice}. This extra cost can refer to time, energy or monetary cost, so three elements are provided: \textit{extraProcessingTime}, \textit{extraEnergyConsumption} and \textit{extraMonetaryCost}, as shown in the following example.

\lstset{language=XML,tabsize=2,
    commentstyle=\color{OliveGreen},
    stringstyle=\color{red},
    numbers=none,
    numberstyle=\tiny,
    numbersep=5pt,
    breaklines=true,
    showstringspaces=false,
    basicstyle=\footnotesize,
    emph={name,availability,source,destination,processingDevice,extraProcessingTime,extraEnergyConsumption,extraMonetaryCost,start,end},
    xleftmargin=0.1cm,
    columns=flexible,
    emphstyle={\color{blue}}
    }
\begin{lstlisting}
<sequenceDependentSetups>
	<sequenceDependentSetup source="commodity1Task1" destination="commodity2Task1" processingDevice="device1">
    	<extraProcessingTime>x1</extraProcessingTime>
    	<extraEnergyConsumption>y1</extraEnergyConsumption>
    	<extraMonetaryCost>z1</extraMonetaryCost>
	</sequenceDependentSetup>
</sequenceDependentSetups>
\end{lstlisting}

\section{Real-World Use Cases} \label{sec:usecase}
In this section, two real-world business cases (related to the discrete and process manufacturing branches) are used to demonstrate the adoption of the proposed Factory Description Language.

\subsection{Discrete Manufacturing Scenario}

The considered discrete manufacturing scenario is related to wire electrical discharge machining (WEDM), where a thermo-electric sparking process removes material using a wire to cut the desired shape of a part. Complex profiles with tight tolerances in hard conductive materials can be obtained. The objectives are to minimise the makespan and the monetary cost per part. This cost can be obtained by summarising all values of \textit{monetaryCost} of the subprocesses involved in producing a given part. If applicable, the values of \textit{extraMonetaryCost} of \textit{sequenceDependentSetups} should be added as well.

The resource allocation consists of selecting processing devices (and thus production line) for cutting the part (product). The selected processing devices (machines) can process parts of various sizes (small, medium or large) and operate in a number of modes, each related to, e.g., a different wire type. Consequently, all possible modes for processing each considered part have to be explicitly specified using element \textit{subprocessProcessingDeviceMode}. Table~\ref{tab:parts} presents an example of parameters of manufacturing parts in the considered factory.

\begin{table}[ht]
\caption{Example parameters of cutting parts in the considered discrete manufacturing scenario}\label{tab:parts}
\resizebox{0.48\textwidth}{!}{%
\begin{tabular}{|c|c|c|c|c|c|c|}
\hline
\thead{Part} & \thead{Machine Size} & \thead{Mode} & \thead{Cutting\\time\\(min)} & \thead{Wire\\Cost\\per \\part\\(\$)}  & \thead{Machine\\cost\\per\\part\\(\$)} & \thead{Total\\cost\\per\\part\\(\$)} \\ \hline
\thead{Part\\01} &

\thead{Small\\Small\\Small\\Small\\Medium\\$\ldots$\\Large} & \thead{1\\2\\3\\4\\1\\$\ldots$\\4} &

\thead{2833.5\\2956.2\\3042.1\\3174.1\\2033.5\\$\ldots$\\1974.1} &

\thead{28.1\\28.1\\28.1\\30.2\\30.2\\$\ldots$\\53.7} &

\thead{164.0\\140.3\\147.8\\136.8\\242.9\\$\ldots$\\408.4} &

\thead{192.1\\168.4\\175.9\\167.0\\273.1\\$\ldots$\\462.1 }
\\\hline

 \multicolumn{6}{c}{$\ldots$} \\ \hline
\thead{Part\\20} & \thead{Large\\Large\\Large\\Large} & \thead{1\\2\\3\\4} & \thead{5341.3\\5505.1\\5191.7\\4106.6} & \thead{335.2\\383.1\\482.1\\648.3} & \thead{5866.7\\8381.0\\5673.8\\4754.9} & \thead{6201.9\\8764.1\\5673.8\\4754.9} \\ \hline

    \end{tabular}}
\end{table}

With FDL, each part in the considered discrete manufacturing scenario is characterised by its name, its priority (in terms of urgency), the number of cuts required to produce the part and the list of compatible devices of the given production. Besides, a production contains a set of subprocesses representing each cut operation, where a production procedure can be pre-empted between cuts. Each subprocess (a cut) contains the information of processing time, energy consumption and monetary cost for executing on a given machine.
Note, the above configuration is built based on the consideration that users may need to configure each cut operation manually (e.g., adjust the processing time). In the case where manually configuring cut operations is not necessary, a user can also describe a production without providing information of subprocesses, where the system generates the corresponding subprocesses automatically based on the given number of cuts.

\begin{figure}[b]
\centering
\includegraphics[width=\columnwidth]{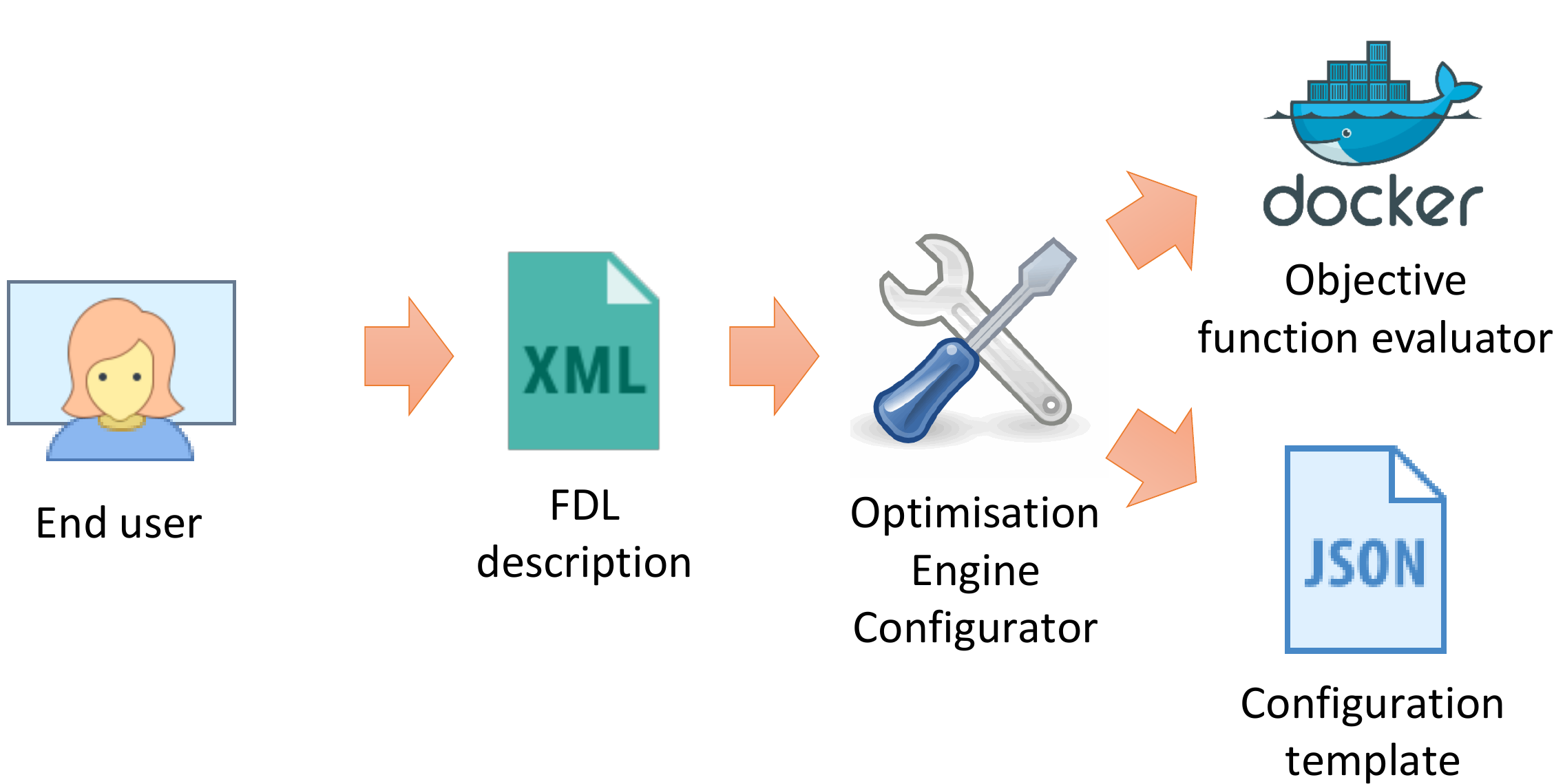}
\caption{Optimisation Engine Configurator use flow}\label{fig:overallArchitecture}
\end{figure}

Then, the Optimisation Engine Configurator (OEC) is used to generate both the configuration template and the objective function evaluator in the following way. Depending on the input factory parameter, OEC locates to the correct XML factory modelling file and reads the corresponding optimisation parameters and factory descriptions, which include optimisation objectives, factory resources with their availability, production processes with their subprocesses, subprocess relations of subprocesses and dependent setups for production processes. This flow is illustrated in Figure \ref{fig:overallArchitecture}. An example of the FDL-based factory model for the considered discrete manufacturing scenario is given below, starting with the objective description.

\lstset{language=XML,tabsize=2,
    commentstyle=\color{OliveGreen},
    stringstyle=\color{red},
    numbers=none,
    numberstyle=\tiny,
    numbersep=5pt,
    breaklines=true,
    showstringspaces=false,
    basicstyle=\footnotesize,
    emph={name,availability,source,destination,processingDevice,extraProcessingTime,extraEnergyConsumption,extraMonetaryCost,start,end},
    xleftmargin=0.1cm,
    columns=flexible,
    emphstyle={\color{blue}}
    }
\begin{lstlisting}
<objectives>
    <objective name="makespan" />
    <objective name="monetary" />
</objectives>
\end{lstlisting}

The objectives are titled as \textit{objective} with a name specifying the metric for optimisation with the assumption of minimisation. For the the considered discrete manufacturing case, two objectives are supported: \textit{makespan} and \textit{monetary}. However, one can configure the objectives to make the optimisation engine to focus only on one of the objectives by simply removing the other objective.

\lstset{language=XML,tabsize=2,
    commentstyle=\color{OliveGreen},
    stringstyle=\color{red},
    numbers=none,
    numberstyle=\tiny,
    numbersep=5pt,
    breaklines=true,
    showstringspaces=false,
    basicstyle=\footnotesize,
    emph={name,availability,source,destination,extraProcessingTime,extraEnergyConsumption,extraMonetaryCost,start,end,unavailableTime},
    xleftmargin=0.1cm,
    columns=flexible,
    emphstyle={\color{blue}}
    }
\begin{lstlisting}
<processingDevices>
    <processingDevice name="Small 1" availability="1">
        <unavailableTimes>
            <unavailableTime>50,100</unavailableTime>
            <unavailableTime>250,300</unavailableTime>
        </unavailableTimes>
    </processingDevice>
    <processingDevice name="Small 2" availability="1">
        <unavailableTimes>
            <unavailableTime>0,20</unavailableTime>
        </unavailableTimes>
    </processingDevice>
    <processingDevice name="Small 3" availability="0">
        <unavailableTimes>
            <unavailableTime>25,30</unavailableTime>
        </unavailableTimes>
    </processingDevice>
</processingDevices>
\end{lstlisting}

The resources in a factory are modelled as a list of \textit{processingDevice}, where each device is specified with a unique name, availability and unavailable times (in the case where the device is available only in certain time intervals). As given in the example, device \textit{Small 1} is unavailable during periods 50-100 and 250-300 (from the starting point of manufacturing, in minutes) while \textit{Small 2} is available during the entire manufacturing process. For the considered process manufacturing scenario, each device is also associated with an operating mode (i.e., \textit{economy}, \textit{standard} or \textit{performance}), which can be switched dynamically during run-time with various costs imposed during manufacturing.

The production model is presented below. This model contains each part to be produced with a set of subprocesses required to produce it. Each subprocess is then modelled by OEC as an individual task associated with a specified resource allocation (among all its compatible resources) and a unique priority for schedule by a fixed priority preemptive scheduler.

\lstset{language=XML,tabsize=2,
    commentstyle=\color{OliveGreen},
    stringstyle=\color{red},
    numbers=none,
    numberstyle=\tiny,
    numbersep=5pt,
    breaklines=true,
    showstringspaces=false,
    basicstyle=\footnotesize,
    emph={name,availability,source,destination,processingDevice,extraProcessingTime,extraEnergyConsumption,extraMonetaryCost,start,end,unavailableTime},
    xleftmargin=0.1cm,
    columns=flexible,
    emphstyle={\color{blue}}
    }
\begin{lstlisting}
<productionProcess name="P15" priority="15" cuts="10">
    <comptiableDevices>
        <comptiableDevice name="Large 1" processingTime="5505" energy="120" montary="4651" />
        <comptiableDevice name="Large 2" processingTime="5341" energy="120" montary="6573" />
        <comptiableDevice name="Large 3" processingTime="7421" energy="120" montary="3566" />
        <comptiableDevice name="Large 4" processingTime="6205" energy="120" montary="4255" />
    </comptiableDevices>
    <subProcesses>
        <subProcess name="P15 cut 1">
          <subProcessProcessingDevice name="Large 1" processingTime="550" energy="12" montary="465" />
          <subProcessProcessingDevice name="Large 2" processingTime="534" energy="12" montary="657" />
          <subProcessProcessingDevice name="Large 3" processingTime="742" energy="12" montary="356" />
          <subProcessProcessingDevice name="Large 4" processingTime="620" energy="12" montary="425" />
        </subProcess>
        <subProcess name="P15 cut 2">
          <subProcessProcessingDevice name="Large 1" processingTime="550" energy="12" montary="465"/>
          <subProcessProcessingDevice name="Large 2" processingTime="534" energy="12" montary="657"/>
          <subProcessProcessingDevice name="Large 3" processingTime="742" energy="12" montary="356"/>
          <subProcessProcessingDevice name="Large 4" processingTime="620" energy="12" montary="425"/>
        </subProcess>
        <subProcess name="P15 cut 3">
          <subProcessProcessingDevice name="Large 1" processingTime="550" energy="12" montary="465" />
          <subProcessProcessingDevice name="Large 2" processingTime="534" energy="12" montary="657" />
          <subProcessProcessingDevice name="Large 3" processingTime="742" energy="12" montary="356" />
          <subProcessProcessingDevice name="Large 4" processingTime="620" energy="12" montary="425" />
        </subProcess>
    </subProcesses>
</productionProcess>
\end{lstlisting}

To ensure that the subprocesses of a production process are executed in the correct order (if necessary), the notion \textit{subprocessRelation} is introduced to describe the execution sequence of subprocesses. For either the considered discrete or process manufacturing scenario, the notion \textit{M} is used to describe that a subprocess \textit{SP2} is executed immediately after another subprocess \textit{SP1}. This information is used by OEC for generating the objective function.

\lstset{language=XML,tabsize=2,
    commentstyle=\color{OliveGreen},
    stringstyle=\color{red},
    numbers=none,
    numberstyle=\tiny,
    numbersep=5pt,
    breaklines=true,
    showstringspaces=false,
    basicstyle=\footnotesize,
    emph={name,availability,source,destination,processingDevice,extraProcessingTime,extraEnergyConsumption,extraMonetaryCost,start,end,unavailableTime},
    xleftmargin=0.1cm,
    columns=flexible,
    emphstyle={\color{blue}}
    }
\begin{lstlisting}
<subprocessRelations>
    <subprocessRelation source="P1 cut 0" destination="P1 cut 1" allensOperator="M" />
    <subprocessRelation source="P1 cut 1" destination="P1 cut 2" allensOperator="M" />
    <subprocessRelation source="P2 cut 0" destination="P2 cut 1" allensOperator="M" />
    <subprocessRelation source="P2 cut 1" destination="P2 cut 2" allensOperator="M" />
    <subprocessRelation source="P3 cut 0" destination="P3 cut 1" allensOperator="M" />
    <subprocessRelation source="P3 cut 1" destination="P3 cut 2" allensOperator="M" />
  </subprocessRelations>
\end{lstlisting}

At last, during scheduling, multiple productions could execute on one resource, which could cause the extra cost for the machine to be cleaned and/or reset for a different product. The FDL models treat such cost as a set of potential independent tasks for sequence-dependent setup, where each of such tasks describes the source production (the product being processed), the target production (the product to be processed), the resource, time consumption and the corresponding costs. The objective function will be generated by OEC based on the list of dependent setup tasks and applies these tasks dynamically where appropriate (i.e., when a dependent setup is necessary) during the scheduling process.

\lstset{language=XML,tabsize=2,
    commentstyle=\color{OliveGreen},
    stringstyle=\color{red},
    numbers=none,
    breaklines=true,
    showstringspaces=false,
    basicstyle=\footnotesize,
    emph={name,availability,source,destination,processingDevice,extraProcessingTime,extraEnergyConsumption,extraMonetaryCost,start,end,unavailableTime},
    xleftmargin=0.1cm,
    columns=flexible,
    emphstyle={\color{blue}}
    }
\begin{lstlisting}
<sequenceDependentSetups>
    <sequenceDependentSetup source="P1" destination="P2" processingDevice="Small 4" extraProcessingTime="10" extraEnergyConsumption="10" extraMonetaryCost="1000" />
    <sequenceDependentSetup source="P1" destination="P2" processingDevice="Medium 1" extraProcessingTime="10" extraEnergyConsumption="10" extraMonetaryCost="1000" />
    <sequenceDependentSetup source="P1" destination="P2" processingDevice="Small 3" extraProcessingTime="10" extraEnergyConsumption="10" extraMonetaryCost="1000" />
    <sequenceDependentSetup source="P1" destination="P2" processingDevice="Medium 2" extraProcessingTime="10" extraEnergyConsumption="10" extraMonetaryCost="1000" />
    <sequenceDependentSetup source="P1" destination="P2" processingDevice="Small 2" extraProcessingTime="10" extraEnergyConsumption="10" extraMonetaryCost="1000" />
    <sequenceDependentSetup source="P1" destination="P2" processingDevice="Small 1" extraProcessingTime="10" extraEnergyConsumption="10" extraMonetaryCost="1000" />
    <sequenceDependentSetup source="P1" destination="P2" processingDevice="Large 3" extraProcessingTime="10" extraEnergyConsumption="10" extraMonetaryCost="1000" />
    <sequenceDependentSetup source="P1" destination="P2" processingDevice="Large 4" extraProcessingTime="10" extraEnergyConsumption="10" extraMonetaryCost="1000" />
    <sequenceDependentSetup source="P1" destination="P2" processingDevice="Medium 3" extraProcessingTime="10" extraEnergyConsumption="10" extraMonetaryCost="1000" />
    <sequenceDependentSetup source="P1" destination="P2" processingDevice="Large 1" extraProcessingTime="10" extraEnergyConsumption="10" extraMonetaryCost="1000" />
    <sequenceDependentSetup source="P1" destination="P2" processingDevice="Medium 4" extraProcessingTime="10" extraEnergyConsumption="10" extraMonetaryCost="1000" />
    <sequenceDependentSetup source="P1" destination="P2" processingDevice="Large 2" extraProcessingTime="10" extraEnergyConsumption="10" extraMonetaryCost="1000" />
</sequenceDependentSetups>
\end{lstlisting}

FDL can be applied to other discrete manufacturing scenarios in a similar manner.

\subsection{Process Manufacturing Scenario}
\begin{table}[b]
\caption{Example of parameter for producing several paints the considered process manufacturing scenario}\label{tab:paints}
\centering
\resizebox{0.4\textwidth}{!}{%
\begin{tabular}{|c|c|c|c|c|}
\hline
\thead{Paint\\Types}	 & 	\thead{Compatible \\Production Lines}	& \thead{Amount of\\Produced\\Commodity}& \thead{Recipe\\Execution\\Time } \\ \hline
\thead{Std\\White}	&	\thead{$\{ P_{1}, P_{2}, P_{3},$  $P_{4}, P_{5} \}$ \\ $\{P_{6}, P_{7} \}$ \\ $\{P_{8}, P_{9} \}$}	&	\thead{5 t \\ 10 t \\ 10 t }	& \thead{	60 min.\\  45 min. \\ 30 min.} \\ \hline
\thead{Super\\White}	&		\thead{$\{ P_{1}, P_{2}, P_{3},$ $P_{4}, P_{5} \}$ \\ $\{P_{6}, P_{7} \}$ \\ $\{P_{8}, P_{9} \}$ }	&	\thead{6 t\\ 12 t \\ 12 t }	&	\thead{	90 min.\\ 60 min. \\ 45 min.} \\ \hline
\thead{Std\\Blue}	&		\thead{$\{ P_{1}, P_{2}, P_{3},$ $P_{4}, P_{5} \}$ \\ $\{P_{6}, P_{7} \}$ \\ $\{P_{8}, P_{9} \}$}	&		\thead{4 t\\  8 t \\ 8 t  }	&	\thead{	100 min.\\80 min. \\ 60 min. } \\ \hline
\thead{Std\\Green}	&			\thead{$\{ P_{1}, P_{2}, P_{3},$ $P_{4}, P_{5} \}$ \\ $\{P_{6}, P_{7} \}$ \\ $\{P_{8}, P_{9} \}$ }	&	\thead{4 t\\ 8 t \\ 8 t }	&\thead{	120 min. \\ 90 min. \\ 60 min. } \\  \hline
    \end{tabular}}
\end{table}

The description of the considered process manufacturing factory is similar to that of the considered discrete manufacturing scenario, but with the notion \textit{productLine} introduced. The reason to introduce this notation is that, as a typical process manufacturing plant, commodities produced often require several devices working in collaboration, with an explicit execution order enforced.

The considered chemical plant produces paints by mixing/dispersion of powdery, liquid and paste recipe components, following a stored recipe. Table~\ref{tab:paints} gives an example of parameters for producing several paints the considered process manufacturing scenario.
Below we provide an example FDL for describing the production line.

\lstset{language=XML,tabsize=2,
    commentstyle=\color{OliveGreen},
    stringstyle=\color{red},
    numbers=none,
    breaklines=true,
    showstringspaces=false,
    basicstyle=\footnotesize,
    emph={name,availability,source,destination,processingDevice,extraProcessingTime,extraEnergyConsumption,extraMonetaryCost,start,end,unavailableTime,order},
    xleftmargin=0.1cm,
    columns=flexible,
    emphstyle={\color{blue}}
    }
\begin{lstlisting}
<productionLines>
<productionLine name="ProductionLine 1">
  <productionLineProcessingDevices>
    <productionLineProcessingDevice order="0" name="Silo 1" />
    <productionLineProcessingDevice order="1" name="Mixer 1" />
    <productionLineProcessingDevice order="2" name="Tank 1" />
  </productionLineProcessingDevices>
</productionLine>
<productionLine name="ProductionLine 2">
  <productionLineProcessingDevices>
    <productionLineProcessingDevice order="0" name="Silo 1" />
    <productionLineProcessingDevice order="1" name="Mixer 2" />
    <productionLineProcessingDevice order="2" name="Tank 1" />
  </productionLineProcessingDevices>
</productionLine>
<productionLine name="ProductionLine 3">
  <productionLineProcessingDevices>
    <productionLineProcessingDevice order="0" name="Silo 2" />
    <productionLineProcessingDevice order="1" name="Mixer 3" />
    <productionLineProcessingDevice order="2" name="Tank 1" />
  </productionLineProcessingDevices>
</productionLine>
</productionLines>
\end{lstlisting}

As given above, the products in the considered process manufacturing scenario are manufactured by production lines, where each line contains certain devices. All lines are executed following a strict execution order: Silo, Mixer and Tank. Below presents the example describing the production process and related-cost for producing the paint \textit{Std Weiss}.

\lstset{language=XML,tabsize=2,
    commentstyle=\color{OliveGreen},
    stringstyle=\color{red},
    numbers=none,
    breaklines=true,
    showstringspaces=false,
    basicstyle=\footnotesize,
    emph={name,availability,source,destination,processingDevice,extraProcessingTime,extraEnergyConsumption,extraMonetaryCost,start,end,unavailableTime},
    xleftmargin=0.1cm,
    columns=flexible,
    emphstyle={\color{blue}}
    }
\begin{lstlisting}
<productionProcess name="Std Weiss 45t" priority="1">

<processType name="Std Weiss A" amountProduced="5t">

<comptiableProductionLines>
  <comptiableProductionLine>ProductionLine 1</comptiableProductionLine>
  <comptiableProductionLine>ProductionLine 2</comptiableProductionLine>
  <comptiableProductionLine>ProductionLine 3</comptiableProductionLine>
</comptiableProductionLines>

<subprocesses>
<subprocess name="Std Weiss A Task 1">
<subprocessProcessingDevicesGroup processingTime="15">
  <subprocessProcessingDevice name="Silo 1" mode="Ecomony"/>
  <subprocessProcessingDevice name="Mixer 1" mode="Ecomony"/>
</subprocessProcessingDevicesGroup>
<subprocessProcessingDevicesGroup processingTime="15">
  <subprocessProcessingDevice name="Silo 1" mode="Ecomony"/>
  <subprocessProcessingDevice name="Mixer 1" mode="Standard"/>
</subprocessProcessingDevicesGroup>
<subprocessProcessingDevicesGroup processingTime="15">
  <subprocessProcessingDevice name="Silo 1" mode="Ecomony"/>
  <subprocessProcessingDevice name="Mixer 1" mode="Power"/>
</subprocessProcessingDevicesGroup>
</subprocess>
<subprocess name="Std Weiss A Task 2">
<subprocessProcessingDevicesGroup processingTime="120">
  <subprocessProcessingDevice name="Mixer 1" mode="Ecomony" />
</subprocessProcessingDevicesGroup>
<subprocessProcessingDevicesGroup processingTime="80">
  <subprocessProcessingDevice name="Mixer 1" mode="Standard" />
</subprocessProcessingDevicesGroup>
<subprocessProcessingDevicesGroup processingTime="40">
   <subprocessProcessingDevice name="Mixer 1" mode="Power" />
</subprocessProcessingDevicesGroup>
<subprocess name="Std Weiss A Task 3">
<subprocessProcessingDevicesGroup processingTime="10">
  <subprocessProcessingDevice name="Mixer 1" mode="Ecomony" />
  <subprocessProcessingDevice name="Tank 1" mode="Standard" />
</subprocessProcessingDevicesGroup>
<subprocessProcessingDevicesGroup processingTime="10">
  <subprocessProcessingDevice name="Mixer 1" mode="Standard" />
  <subprocessProcessingDevice name="Tank 1" mode="Standard" />
</subprocessProcessingDevicesGroup>
<subprocessProcessingDevicesGroup processingTime="10">
  <subprocessProcessingDevice name="Mixer 1" mode="Power " />
  <subprocessProcessingDevice name="Tank 1" mode="Standard" />
</subprocessProcessingDevicesGroup>
\end{lstlisting}

As described in the FDL extract above, an order of 45 tonnes of \textit{Std Weiss} is executed by several sub-orders, where each sub-order produces the maximum amount that the production line can produce in one execution e.g., \textit{Std Weiss A} can produce 5 tonnes in each execution on production line $P_1$.

\lstset{language=XML,tabsize=2,
    commentstyle=\color{OliveGreen},
    stringstyle=\color{red},
    numbers=none,
    breaklines=true,
    showstringspaces=false,
    basicstyle=\footnotesize,
    emph={name,availability,source,destination,processingDevice,extraProcessingTime,extraEnergyConsumption,extraMonetaryCost,start,end,unavailableTime},
    xleftmargin=0.1cm,
    columns=flexible,
    emphstyle={\color{blue}}
    }
\begin{lstlisting}
<subprocessRelations>
    <subprocessRelation source="Std Weiss A Task 1" destination="Std Weiss A Task 2" allensOperator="M" />
    <subprocessRelation source="Std Weiss A Task 2" destination="Std Weiss A Task 3" allensOperator="M" />
</subprocessRelations>
\end{lstlisting}

The Element \textit{subprocessRelations} contains a set of sub-elements describing the execution order between subprocesses in a given process, where  Allen's temporary operator \textit{M} indicates that the source subprocess must be executed directly before the destination.

FDL can be applied to other process manufacturing scenarios in a similar manner.


\section{Conclusion}\label{sec:conclusion}

In this paper, a Factory Description Language for describing manufacturing plants is presented to facilitate the use of re-configurable planning and scheduling. The proposed FDL deliver critical information towards optimisation, required production and manufacturing process that will be consumed by an optimisation engine running in the cloud. With the FDL elements presented in this paper, the optimisation objectives, required products and the production processes of both the considered discrete and process manufacturing cases are effectively described by the proposed FDL. In addition, the FDL is formed in a human-readable fashion and can be easily altered by engineers who do not have high-level expertise of the related optimisation systems. With the proposed FDL, we provide an approach that describes the input information required by cloud-based optimisations engines for dynamic planning and scheduling in smart factories.

The FDL provides the fundamental approach to feed the input information to the cloud-based optimisation engine proposed in previous works for re-configuring manufacturing plants for improved profits with minimised makespan. Two real-world use case scenarios (with two mainstream manufacturing types) are provided to illustrate the factory description based on the proposed FDL. In future, this factory description and modelling approach will be extended to support a wider range of business cases, such as different types of manufacturing plants and smart restaurants.

\section*{Acknowledgement}
The authors acknowledge the support of the EU H2020 SAFIRE project
(Ref. 723634).

The icons used in this paper have been created by Icons8, https://icons8.com.


\balance

\bibliographystyle{IEEEtran}
\bibliography{fdl}


\end{document}